\documentclass[10pt,aps,prl,groupedaddress,twocolumn,amsmath,amssymb,showpacs,floatfix]{revtex4-1}
\usepackage[latin1]{inputenc}

\usepackage{amsmath}
\usepackage{amsfonts}
\usepackage{amssymb}
\usepackage{natbib}
\usepackage{float}
\usepackage{color}
\usepackage{graphicx}% Include figure files
\usepackage{dcolumn}% Align table columns on decimal point
\usepackage{bm}% bold math
\usepackage{pinlabel}
\usepackage{braket}

\newcommand{\LibraryFile}{/Users/achaywa/Documents/physics/research/library}
\newcommand{\etal}{{\it et al. }}

\newcommand{\beq}{\begin{equation}}
\newcommand{\eeq}{\end{equation}}
\newcommand{\beqarray}{\begin{eqnarray}}
\newcommand{\eeqarray}{\end{eqnarray}}

\newcommand{\eqn}[2][]{\begin{equation} #2 #1\end{equation}}
\newcommand{\en}[1]{\ensuremath{#1}}
\newcommand{\fig}[2]{\begin{figure} #1 \end{figure}}
\newcommand{\eqna}[2][]{\eqn{\begin{array}{#1} #2 \end{array}}}
\newcommand{\symm}{\en{\mathcal{S}}}
\newcommand{\oneOn}[1]{\en{\frac{1}{#1}}}
\newcommand{\AtomicGroundState}{\en{g}}
\newcommand{\AtomicExcitedState}[1][]{\en{e_{#1}}}
\newcommand{\CavityPhotonNumber}{\en{n}}
\newcommand{\JCBareState}[2]{\en{|#1 , #2\rangle}}
\newcommand{\JaynesCummingsHamiltonian}[1][]{\en{H^{JC}_{#1}}} 
\newcommand{\CavityDetuning}[1][]{\en{\Delta_{#1}}}
\newcommand{\CavityFrequency}[1][]{\en{\omega_{#1}}}
\newcommand{\CavityExitationNumber}[1][]{\en{\ell_{#1}}}
\newcommand{\CavityExitationNumberOperator}[1][]{\en{L_{#1}}}
\newcommand{\AtomicEnergy}[1][]{\en{\epsilon_{#1}}}
\newcommand{\AtomCavityCoupling}[1][]{\en{\beta_{#1}}}
\newcommand{\JaynesCummingsHamiltonianDefinition}{\en{\CavityFrequency \CavityExitationNumberOperator + \CavityDetuning\sigma^+\sigma^- + \AtomCavityCoupling (\sigma^{+}a+\sigma^{-}a^{\dagger})}} 
\newcommand{\JaynesCummingsHamiltonianEquation}[1][]{\eqn[#1]{\JaynesCummingsHamiltonian = \JaynesCummingsHamiltonianDefinition \label{JaynesCummingsHamiltonianEquation} }} 

\newcommand{\TunnelingRate}[1][]{\en{\kappa_{#1}}}
\newcommand{\NearestNeighbors}[2]{\en{\langle{}#1,#2\rangle}}
\newcommand{\HubbardHamiltonianDefinition}{\en{-\sum\limits_{\NearestNeighbors{i}{j}}\TunnelingRate[ij]a^\dagger_ia_j}}

\newcommand{\JaynesCummingsHubbardHamiltonian}{\en{H^{JCH}}} 
\newcommand{\JCHDefinition}{\en{\sum\limits_i\JaynesCummingsHamiltonian[i] \HubbardHamiltonianDefinition}}
\newcommand{\JCHHamiltonianEquation}[1][]{\eqn[#1]{\JaynesCummingsHubbardHamiltonian =  \JCHDefinition}}

\newcommand{\fillingFactor}{\en{\nu}}
\newcommand{\numberOfParticles}{\en{N_p}}

\newcommand{\ChernNumber}{\en{\mathcal{C}}}

\newcommand{\threeBodyBHStrength}{\en{U_3}}
\newcommand{\twoBodyBHStrength}{\en{U_2}}
\newcommand{\threeLevelConfiguration}{\en{\Xi}}
\newcommand{\GeneralizedPeriodicBoundaryConditions}[1][]{\en{\theta_{#1}}}

\newcommand{\ThreeLevelAtomHamiltonianEquationP}{
    \begin{equation}
        \begin{split}
        H^{3L} = \CavityFrequency a^\dagger a & + \AtomicEnergy[1]\ket{e_1}\bra{e_1} + \left(\ThreeLevelAtomCouplingOne \sigma^{+}_{1}a + h.c. \right) \\
       & + \AtomicEnergy[1]\ket{e_2}\bra{e_2} + \left(\ThreeLevelAtomCouplingTwo \sigma^{+}_{2}a + h.c. \right).
    \end{split}
        \end{equation}}

\newcommand{\ThreeLevelAtomCouplingOne}{\en{\beta_{1}}}
\newcommand{\ThreeLevelAtomCouplingTwo}{\en{\beta_{2}}}
\newcommand{\threeLevelAtomEnergies}[1][]{\eqn[#1]{\AtomicEnergy[1] = \omega-\Delta,\qquad  \AtomicEnergy[2]= 2\CavityFrequency - 2\CavityDetuning}}
\newcommand{\threeLevelAtomStrengths}[1][]{\eqn[#1]{g\leftrightarrow e_1 \equiv \ThreeLevelAtomCouplingOne,\qquad e_1\leftrightarrow e_2 \equiv \ThreeLevelAtomCouplingTwo = \sqrt{2}\ThreeLevelAtomCouplingOne}}
\newcommand{\magneticLength}{\en{\ell_B}}
\newcommand{\LLLcoordinates}[1][]{\en{z_{#1} = \left(x_{#1} + iy_{#1}\right)/\magneticLength}}

\newcommand{\JastrowFunction}[1][]{\en{F_{#1}\left(\coordinateVector\right)}}
\newcommand{\QHEGaussianFactor}{\en{\exp{\left[-\oneOn{4}\sum_{k}|z_k|^2\right]}}}
\newcommand{\LaughlinJastrow}[1][q]{\en{\prod_{k > j}(z_k-z_j)^{#1}}}
\newcommand{\GeneralizedThetaSymbol}{\en{\theta}}
\newcommand{\GeneralizedThetaFunction}[4]{\en{\GeneralizedThetaSymbol
        \left[\begin{array}{c} #1 \\ #2 \end{array}\right]
        \left(#3|#4\right)}}
\newcommand{\GeneralizedThetaForm}{\en{\GeneralizedThetaFunction{a}{b}{z}{\tau}}}
\newcommand{\ThetaFunction}[3]{\en{\theta_{#1}\left(#2|#3\right)}}
\newcommand{\ThetaFunctionForm}[1]{\en{\ThetaFunction{#1}{z}{\tau}}}
\newcommand{\centreOfMassCoordinate}{\en{Z}}
\newcommand{\coordinateVector}{\en{\bar{z}}}
\newcommand{\centreOfMassEquation}{\en{\centreOfMassCoordinate=\sum^{N_p}_kz_k}}
\newcommand{\LLLwavefunction}[1]{\en{\psi^{L}_{#1}}}
\newcommand{\LLLwavefunctionDef}[1]{\en{e^{-y_{#1}^2/4}}}

\newcommand{\CentreOfMassWavefunction}{\en{F_{cm}(\centreOfMassCoordinate)}}
\newcommand{\CentreOfMassWavefunctionDef}{\GeneralizedThetaFunction{N_p/q +(N_\phi{} -2)/2q}{-(N_\phi -2)/q}{q\frac{Z}{L_x}}{iq\frac{L_y}{L_x}}}
\newcommand{\relativeWavefunction}{\en{f_{rel}(\coordinateVector)}}

\newcommand{\relativeWavefunctionDef}{\en{\prod\limits_{i<j}^{\numberOfParticles}\ThetaFunction{1}{\frac{z_i - z_j}{L_x}}{i\frac{L_y}{L_x}}^q}}

\newcommand{\LaughlinTorus}[1]{\en{\Psi_L(#1)}}
\newcommand{\LaughlinWavefunctionTorus}{\en{\CentreOfMassWavefunction\relativeWavefunction\prod\limits_{i}^{\numberOfParticles}\LLLwavefunction}}
\newcommand{\LaughlinTorusEquation}[1][]{\eqn[#1]{\LaughlinTorus{\coordinateVector} \propto \LaughlinWavefunctionTorus}}
\newcommand{\LaughlinTorusDefinitionEquations}[1][]{\eqna[rcl]{\CentreOfMassWavefunction & = & \CentreOfMassWavefunctionDef , \\
                                                        \relativeWavefunction & = & \relativeWavefunctionDef #1}}

\newcommand{\pfaffianWavefunction}[1][q]{\mathrm{Pf}\left(\frac{1}{z_k-z_j}\right)\LaughlinJastrow[#1]}
\newcommand{\PfaffSymbol}{\en{\Psi^\mathrm{Pf}}}
\newcommand{\PfaffEquation}[1][]{\eqn[#1]{\label{PfaffEquation} \PfaffSymbol \propto \pfaffianWavefunction}}
\newcommand{\PfaffDoubleLaughlinDef}{\en{\symm\LaughlinTorus{z_{1,\ldots,\numberOfParticles}}
        \LaughlinTorus{z_{\ldots\numberOfParticles+1,\dots,\numberOfParticles}}}}
\newcommand{\PfaffDoubleLaughlinDefEqn}[1][]{\eqn[#1]{\PfaffSymbol = \PfaffDoubleLaughlinDef}}

\newcommand{\threeLevelAtomFigure}{\begin{figure*}[t]
\begin{tabular}{lccr}
    \hspace{-0cm}
    \begin{minipage}{2.5cm}    
    \labellist
        \small\hair 2pt
    \pinlabel $a)$ at -10 325
    \endlabellist
    $\vtop{\hbox{\includegraphics[scale=0.4]{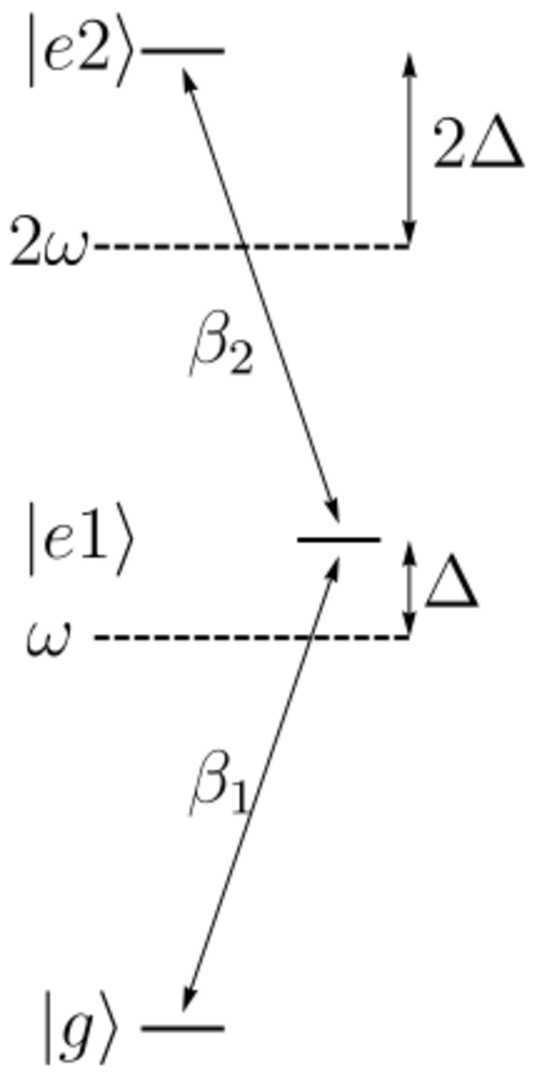}}}$
\end{minipage} &
\begin{minipage}{6.2cm}    
    \labellist
        \small\hair 2pt
    \pinlabel $b)$ at 10 130
    \endlabellist
    \includegraphics[scale=1]{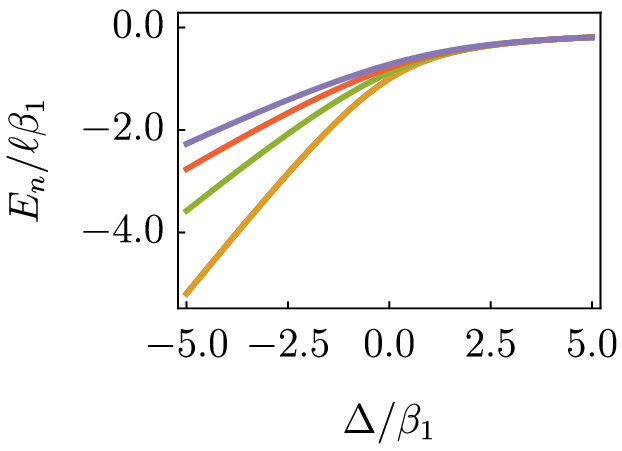}
\end{minipage} &
\begin{minipage}{6.0cm}    

    \labellist
        \small\hair 2pt
    \pinlabel $c)$ at 10 130
    \endlabellist
    \includegraphics[scale=1]{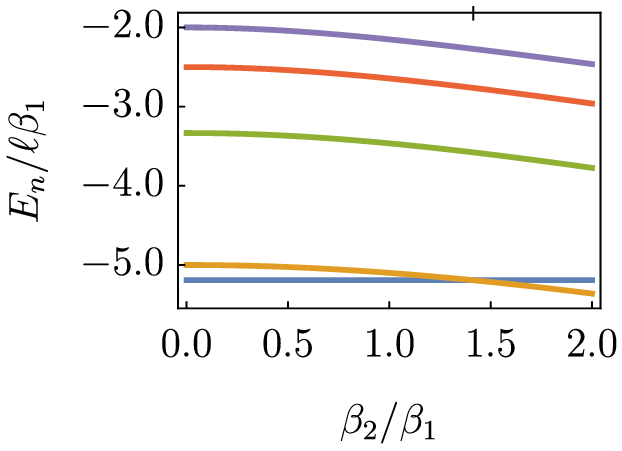}
\end{minipage}
\begin{minipage}{1.5cm} 
    \vspace{-0.5cm}
    \includegraphics[scale=1]{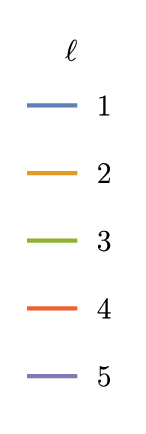}
\end{minipage}
\end{tabular}
\caption{a) 3 level atom configuration for inducing a 3-body interaction in the JCH
lattice. b) Energy per excitation in the 3-level atom for $n=1-5$(blue, orange, green red, purple in order). Excitations
in the $n=1$ and $n=2$ excitation subspace have the same energy cost. For
higher excitation numbers, the energy per particle increases. c) Energy per excitation as a function of \AtomCavityCoupling[2] (same colors as b) for $\CavityDetuning/\AtomCavityCoupling[1] = -5$. \label{threeLevelAtomFigure}}
\vspace{-0.5cm}
\end{figure*}}

\newcommand{\BothPfaffianBandgapFigure}{\fig{

   \begin{center}
       \includegraphics[scale=0.4]{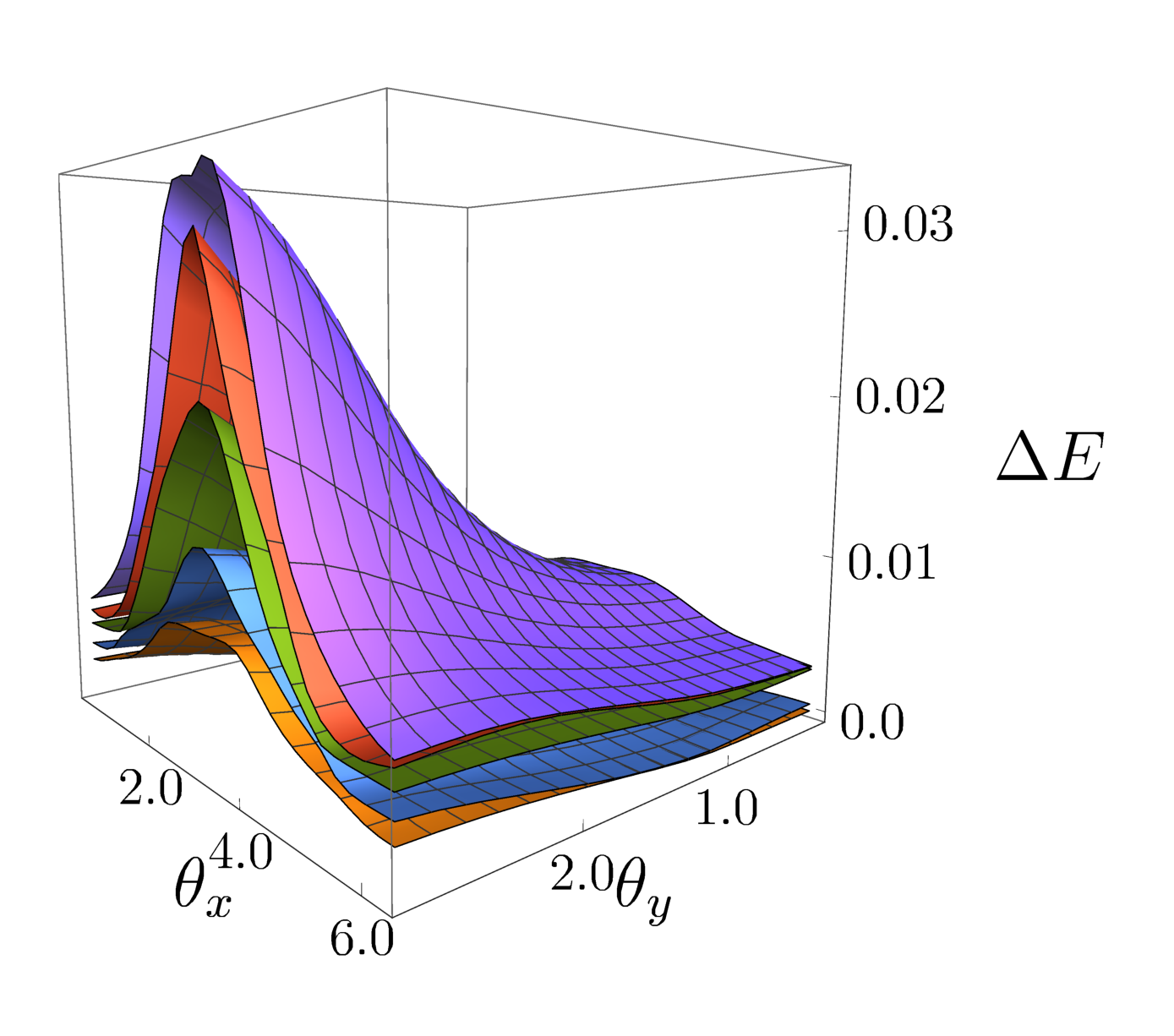}
 \end{center}
 \vspace{-1cm}
 \caption{
Energy gap for the first 5 excited states in the 3-level JCH model on a
$4\times 4$ lattice with 4 particles and 4 flux quanta ($\nu = 1$). There is a 3 dimensional quasi-degeneracy in the
groundstate, indicative of a Pfaffian like state.
   \label{BothPfaffianBandgapFigure}}}}

\newcommand{\PfaffianOverlapFigure}{\fig{
    \labellist
        \small\hair 2pt
    \pinlabel $a)$ at 60 215
    \pinlabel $b)$ at 150 215
    \pinlabel $c)$ at 60 120
    \pinlabel $d)$ at 225 120
    \endlabellist
\centering
\vspace{0cm}\includegraphics[scale=1]{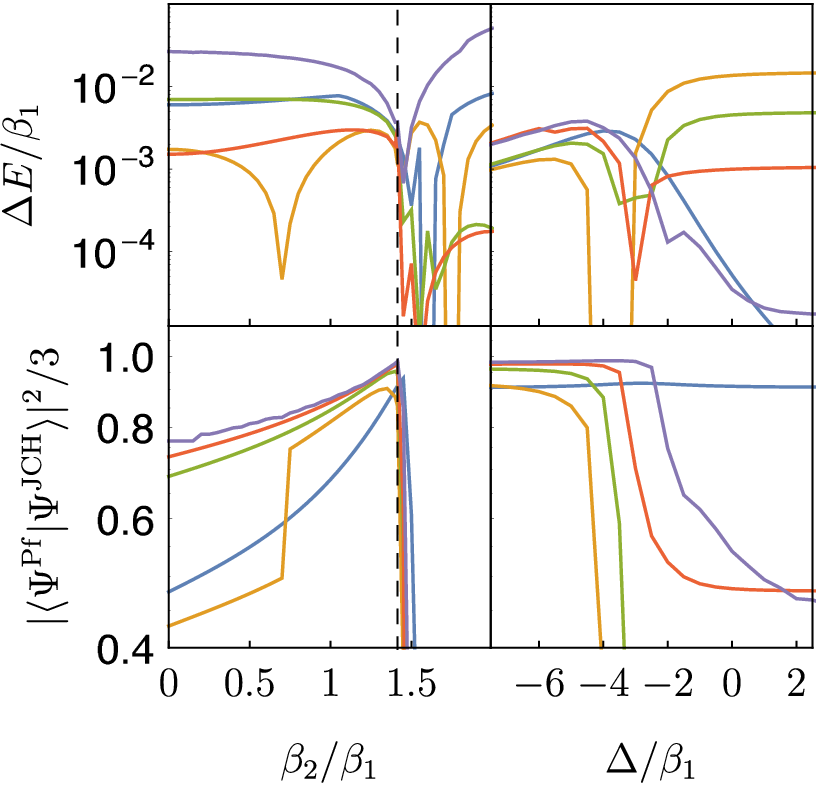} \vspace{0cm}
\caption{
    Bandgap and Pfaffian overlap as a function of \AtomCavityCoupling[2]/\AtomCavityCoupling[1] (at $\CavityDetuning{}/\AtomCavityCoupling[1] = -5$) and
    \CavityDetuning{}/\AtomCavityCoupling[1] (with $\AtomCavityCoupling[2]/\AtomCavityCoupling[1]=\sqrt{2}$) for 4 particles on 4x4 (blue) 4x5 (orange)  5x5 (green) 6x4 (red) 6x5
(purple) and 6x6 (brown) lattice. \label{PfaffianOverlapFigure}}
\vspace{-0.5cm}
}} 

\begin{document}
\title{Pfaffian States in Coupled Atom-Cavity Systems}
\author{Andrew L.C. Hayward}
\affiliation{School of Physics, University of Melbourne, Vic 3050, Australia}
\author{Andrew M. Martin}
\affiliation{School of Physics, University of Melbourne, Vic 3050, Australia}
\date{\today}
\begin{abstract} 

{Coupled atom-cavity arrays, such as those described by the
Jaynes-Cummings Hubbard model, have the potential to emulate a wide range of
condensed matter phenomena. In particular, the strongly correlated states of
the fractional quantum Hall effect can be realised.  At some filling
fractions, the fraction quantum Hall effect has been shown to possess ground
states with non-abelian excitations. The most well studied of these states is
the Pfaffian state of Moore and Read, which is the groundstate of a Hall
Liquid with a 3-body interaction.  In this paper we show how an effective
3-body interaction can be generated within the Cavity QED framework, and that 
a Pfaffian-like groundstate of these systems exists.}\end{abstract}
\maketitle
{%Introduction
    {% motivation : highly correlated states of light. 

Coupled atom-cavity systems (falling under the broad umbrella of Cavity QED) is
proving to be an excellent framework for the investigation of fundamental
quantum phenomena, due to the versatility and control of system parameters. In
particular, cavity QED is promising to be a powerful platform for quantum
emulation, where a wide range of condensed matter systems can be modeled
through tailoring of atom-cavity interactions\cite{Greentree2006,Rossini2007,Quach2009,Gerace2009,Quach2011a,Kapit2014}. }
%TODO citations for this

    {
%A major goal of the quantum emulation scene is to reproduce the physics found
%in electronic systems in other contexts that allow more refined manipulation
%and inspection. Cavity-Atom systems do not have long range interactions, and
%demonstration of a Pfaffian state resulting from purly point iteractions has
%not been demonstrated. 
}

    {% Introduce Pfaffian and motivation 
 
Recent work has shown how complex states of correlated light can be induced in
cavity QED systems via non-linearities in atom-light interactions.  In
particular, a class of states that correspond to the quantum Hall effect have
been predicted. In this paper we extend this class to include Pfaffian-like
states\cite{Moore1991}. These states possess highly non-trivial topological
properties, that make them of great interest to the understanding of quantum
Hall physics, as well as providing an opportunity to examine  strongly
correlated quantum states of light.  }

%These states are produced by inducing an effective 3-body
%interaction.  We first discuss our method for inducing the 3-body interactions
%and derive the effective potential. We then show that the full system has a
%ground state which is topological equivalent to the Pfaffian state for
%appropriate parameters. We discuss the range over which our system will have a
%Pfaffian groundstate by considering the effective interaction. } }

{%Discuss the Pfaffian state, special properties etc, 3 body interaction
Since the discovery of fractional quantum Hall states\cite{Tsui1982}, a
small industry has grown up around the construction of complex states that
might be found in 2-dimensional quantum Hall fluids. These states (for the most
part) lie in the massively degenerate lowest Landau level (LLL). Single particle
states in the lowest Landau level have a beautiful property that leads to the
general structure found across all the QHE states. One can write the position of a
single particle $k$ in the 2D plane as \LLLcoordinates[k], with \magneticLength{} 
the magnetic length, then the single particle states can all be written as a
holomorphic function in $z_k$, multiplied by a Gaussian factor. As a
consequence, all many-body states that lie in the LLL are functions of
$\coordinateVector = \{z_k|k\le\numberOfParticles\}$, and  consist of a Gaussian
term, \QHEGaussianFactor, multiplied by a Jastrow Factor, an analytic
function \JastrowFunction[].

The Jastrow factor encodes the correlations between particles. The simplest
correlated state was proposed by Laughlin\cite{Laughlin1983} to account for
the fractional QHE states found at $\fillingFactor = 1/q$. The Laughlin Jastrow
factors are $\JastrowFunction[L] =   \LaughlinJastrow$. 
%\pdfcomment{Is it Jastrow factor or Jastrow function?} 
These states have a wavefunction that
vanishes as any two particles apporach one another, minimizing the
electron-electron repulsion within the lowest Landau level. 

The Moore-Read Pfaffian state is defined by the Jastrow factor\PfaffEquation[.]
The terms in the Pfaffain factor cancel a factor in the Laughlin state, leading
to a non-zero amplitude at the coincidence of just two particles.

    {%Something about pairing particles
    }

This ground state has some interesting properties; most notably,
non-abelian excitations, and (relatedly) a triply degenerate groundstate in the
toroidal geometry\cite{Moore1991}. }

{%paper summary

    In this work we consider a Jaynes-Cummings-Hubbard (JCH) system with synthetic magnetic field and
demonstrate the existence of Pfaffian-like groundstates at a filling factor
$\fillingFactor = 1$. We first consider the conventional JCH system, and find no evidence for
Pfaffian-like states. We then consider a three level Jaynes-Cummings (JC) cavity and show how
there are regimes in which 3-body interactions dominate. We then study the
conditions under which Pfaffian-like states will arise when these 3 level
systems are Hubbard-coupled. }

{%Introduce the JCH Model

Each cavity in the JCH lattice is described by: \JaynesCummingsHamiltonianEquation[,] where $a$ is the photonic
annihilation operator, $\sigma^{\pm}$ are the atomic raising and lowering
operators, \CavityExitationNumberOperator{} is the excitation number operator,
\CavityDetuning{} the atom-photon detuning, \AtomCavityCoupling{}
the coupling energy and $\hbar=1$. The states \JCBareState{\AtomicGroundState
(\AtomicExcitedState)}{\CavityPhotonNumber }, where \CavityPhotonNumber{} is the
number of photons, and \AtomicGroundState(\AtomicExcitedState) are the ground
(excited) state of the atom, form the single cavity basis.
\JaynesCummingsHamiltonian{} commutes with the total excitation number
operator, \CavityExitationNumberOperator. Therefore the total excitations in
the cavity, \CavityExitationNumber, is a good quantum number.  The eigenstates
of Eq.~(\ref{JaynesCummingsHamiltonianEquation}) are termed polaritons,
superpositions of atomic and photonic excitations, and are a function of
\CavityExitationNumber{} and \CavityDetuning/\AtomCavityCoupling. 
   
The JCH model describes an array of individual Jaynes-Cummings cavities,
which are coupled via a Hubbard like photon tunneling term. In the case of a
lossless system, the JCH model can be described by:
\JCHHamiltonianEquation[,] 
where \TunnelingRate[ij] is the tunneling rate between cavities $i$ and $j$ and the
sum over \NearestNeighbors{i}{j} is between nearest neighbors only.

For large detuning ($|\CavityDetuning| \gg \AtomCavityCoupling$), eigenstates
separate out into either atomic or photonic modes.  In this limit, the photonic
or atomic mode can be adiabatically eliminated.  Eliminating the atomic modes,
the photonic mode has a weak Kerr-type photon-photon repulsion\cite{Na2008} and
the exchange of energy between atomic and photonic modes is strongly
suppressed.  However, virtual processes lead to effective interactions in the
photonic and atomic submanifolds.  Photons have an atomic mediated non-linear
onsite repulsion, making  the JCH  model equivalent to the Bose-Hubbard (BH)
model \cite{Hohenadler2011}.  Atomic modes are coupled with the effective
hopping rate
$\TunnelingRate[ij]^{eff}=\TunnelingRate[ij]\AtomCavityCoupling^2/\CavityDetuning^2$\cite{Makin2009}.
As the atomic modes are restricted to two levels, this is effectively a
hardcore boson field for atomic states, in contrast to the weakly-interacting
photon field.}

{% Synthetic magnetic field

Investigation of quantum Hall physics in the JCH model requires the introducion
of a synthetic magnetic field.  An artificial magnetic field may  be realized
via the introduction of some time reversal symmetry breaking interaction. A
number of techniques have been proposed to achieve
this\cite{Haldane2008,Cho2008a,Koch2010a,Kolovsky2011,Umucalilar2011,Kapit2014}.
For example, one may exploit a time-dependent potential to induce magnetic flux
across the lattice\cite{Kolovsky2011}, thereby explicitly breaking the time
symmetry of the system. A similar strategy is proposed in Ref.
\cite{Kapit2014}, where the authors utilize a time dependent inter-site
coupling to induce a synthetic magnetic field.  Alternatively, effective
coupling to real magnetic fields can be used, such as in the proposal by Koch
\etal \cite{Koch2010a}. Other means, such as the use of optically polarized
media\cite{Haldane2008} or via atomically mediated inter-site
coupling\cite{Cho2008a} have also been proposed.

{%Search for Pfaffian in the JCH model. 
Inspired by results in ultra cold atom
simulations\cite{Cooper2001,Regnault2004,Chang2005,Peterson2008}, one might
expect to find evidence of a Pfaffian groundstate in the JCH model at
$\fillingFactor=1$. We numerically investigated the JCH model on a torus to
this end. Existence of a Pfaffian like state can be indicated by a number of
properties of the groundstate: a triply degenerate groundstate manifold,
large overlap with the Pfaffian trial wavefunction, and a Chern number of 3
computed for the three groundstates. 

We conducted a comprehensive search over several parameters within the JCH
model, but the tell-tail signature of a triple degeneracy groundstate proved
elusive. Instead, simulations reveal that for some lattice configurations a
single separated groundstate in the strongly interacting limit, with a
transition to a gapless phase as the effective 2-body interaction decreased.
Other configurations possessed a gapless groundstate extending all the way to
the hardcore limit. While these results jar with the Bose-Einstein condensate (BEC) findings, other
lattice boson simulations have failed similarly\cite{Mazza2010}. 

Although the evidence from BECs suggests that a Pfaffian groundstate should be
preferred, there are other possible states at $\fillingFactor=1$ filling for
Bosons which are in competition with the Pfaffian state. For example,
Read\cite{Read1998} proposes a groundstate in which, approximately, a single
vortex is attached to each Boson. This assignment exactly cancels out the
external magnetic field, which reduces  the problem to that of a Fermi-Liquid.
Alternatively, it is conjectured\cite{Rezayi1999} that a striped phase with
charge density order  may exist at $\fillingFactor= 1$, with some numerical
simulations\cite{Chung2008} finding evidence for this. 

The relativly small size of the systems we have simulated make it difficult to
tease out the importance of different effects which determine the real nature
of the groundstate at this filling factor. However, the poor scaling of these
systems means that significantly larger systems are impractical at this point
in time. Of course, this problem is one of the primary motivators of work into
quantum emulation.

%The Pfaffian state requires a large amount of Landau level degeneracy for the
%pairing of bosons to occur, which may be significantly violated when the
%particles are confined  to a lattice, leading to a non-correlated groundstate
%as suggested.\pdfcomment{This sentence is poor.} Such a groundstate would
%explain the presence of gapless states for some of the simulations at
%$\fillingFactor=1$, although it is not clear how to verify this conjecture at
%this time; It would require the simulations very large systems to separate out
%the impact of  various factors such as finite size, and lattice potentials and
%limited particle number.}

{%Three Body Interaction
    {%We didn't find the Pfaffian state. But it is the groundstate of a three body Hamiltonian

For $\fillingFactor = 1$, the Pfaffian groundstate is the highest density
groundstate of the 3-body delta potential Hamiltonian\cite{Greiter1992}. For
each pair of particles, there is a number of terms in Eq.~(\ref{PfaffEquation}) for
which each particle is in a different partition. However, there is no  term for
which 3 or more particles coincide that does not vanish.  } }

{%Describe the Three Level Atom
    {%explain structure of JCH interaction in terms of photon-photon interactions

The JCH in the limit of large detuning can be described by a Bose-Hubbard with
an effective two body interaction, \twoBodyBHStrength. However, the atomic-cavity
interaction induces interactions to all orders of n-photon interactions.
These higher order interactions are much smaller than the two body interaction,
and therefore the physics of the JCH very much mirrors that of the Bose-Hubbard
model with two body interaction. We now show that it is possible to
eliminate the two-body interaction while retaining the higher order
interactions. }

Three-body interactions (without corresponding 2-body ones) are unnatural, and do
not arise in many physical systems. A number of schemes for creating effective
3-body interactions have been proposed in the context of BECs\cite{Buchler2007,Daley2009,Mazza2010,Daley2014} and in the circuit QED
setting\cite{Hafezi2014}. Below we show that an effective 3-body
interaction can be induced in atom-cavity lattice by replacing the 2-level
atoms in the JCH model with  appropriately tuned 3-level atoms.  Furthermore,
it is demonstrated via simulation that an atom-cavity lattice consisting of
these 3-level atoms can possess a Pfaffian like state as its groundstate.

\threeLevelAtomFigure

We consider the 3 level system atom in the \threeLevelConfiguration{}
configuration, as shown in Fig.~\ref{threeLevelAtomFigure}a). This
configuration consists of two evenly spaced excited states, with the atom
cavity system described by the Hamiltonian:
\ThreeLevelAtomHamiltonianEquationP{} Here, $\sigma^+_{1(2)}$ raises the
atomic level from $g\leftrightarrow e_1$ ($e_1\leftrightarrow e_2$), and levels 1(2)
have energies $\epsilon_{1(2)}$.

Choosing an atom with energy levels: \threeLevelAtomEnergies[,] and transition
strengths: \threeLevelAtomStrengths[,] leads to an effective 3-body
interaction. This can be seen by considering the formulation of the JC system
as a two-mode Bosonic system with inter-mode tunneling. If one imposes a
hardcore boson condition on one of the modes, then the system corresponds
exactly to the JC cavity with a two-level atom. In the single excitation
subspace, the hardcore condition is automatically satisfied, and the system is
simply a free boson model.

If, instead of the hardcore condition, one imposes a 3-particle hardcore
condition ($\threeBodyBHStrength \rightarrow \infty$), then a similar situation
arises, except that for both the single and double excitation subspaces, there
are no interactions. Mapping this model back to the atomic model, for 1 and 2 particles, the equivalent atom-cavity model corresponds
exactly to the one presented previously, where the factor of $\sqrt{2}$ arises
from the indistinguishably of the bosons.

The 3-body non-linearity is demonstrated in Fig.~\ref{threeLevelAtomFigure}
b). Here, the energy cost per particle for the lower polariton branch is
plotted as a function of the detuning, \CavityDetuning{}. For 1 and 2 excitations, the energy per particle is the
same. However, for $\CavityExitationNumber =3$ and above, there is an
increased cost for adding additional particles.
Figure~\ref{threeLevelAtomFigure}c) shows that, at
$\ThreeLevelAtomCouplingTwo/\ThreeLevelAtomCouplingOne = \sqrt{2}$
%
%\pdfcomment{Do i need to show two level atom energies? also, the three level
%atom energies are hard to see. still need to label this figure correctly. Also,
%could have a plot that interpolates between U2=0 and not.}
%
the 2 and 3 level atom-cavity systems share several properties. The
non-linearity is unbounded as the detuning is lowered, and disappears as the
detuning is increased. Also, the non-linearity does not grow quadratically with
excitation number, as is the case for a pure 3-body interaction, similar to the
2-level atom case. 

Figure~\ref{threeLevelAtomFigure}c) demonstrates how the two body interaction is
affected by the strength of \AtomCavityCoupling[2]. As $\beta_2$ is tuned
away from $ \AtomCavityCoupling[2]/\AtomCavityCoupling[1] = \sqrt{2}$ the effective two-body interaction becomes +ve or -ve.
However, at low detunings ($\CavityDetuning/\AtomCavityCoupling[1] = -5$ in Fig.
\ref{threeLevelAtomFigure}c) the relative strength of this effective two body
interaction is much smaller than the three body one. 

This method for generating 3-body interactions opens up unique possibilities
for investigating the physics of topological quantum states that has proved
elusive in traditional environments.  Furthermore, this same technique can be
extended to higher order interactions. There is a hierarchy of states that
generalize the Pfaffian\cite{Read1999} state,  which are expected to be
groundstates of these higher order interactions. The JCH model with this
modification is, to our knowledge, the only system in which such higher order
interactions might be achieved, outside of a fully functional quantum computer.

{%Address stability of 3 level system

In practice, engineering the three level system as described lies well within
the capabilities of current cavity QED fabrication techniques.  Engineering a
system like this in circuit QED has been discussed in \cite{Hafezi2014}. For
cavity atoms, most \threeLevelConfiguration{} configurations tend to be
unstable, with fast relaxation rates that would preclude large scale coherence
in the system. This instability can be mitigated by instead using an $M$ like
configuration (as in \cite{Greentree2000}), where classical driving can be used
to create an effective three level JCH system. } }

{%Demonstrate Pfaffian State using Chern numbers
 
    {% Studied on the torus with PBC To study the groundstate of our 3-level

In our simulations we restrict the system to a torus, to remove edge effects. The toroidal geometry permits twisted periodic
boundary conditions which reduces finite size effects, and allows for
computation of the Chern number. }

    {%Define and motivate Chern Number

The Chern number\cite{Hafezi2007}, \ChernNumber, is a measure of the topology
of the groundstate of the system and can provide evidence for the existence of
a Pfaffian like state. Here, we define a two dimensional manifold over the two
phases, \GeneralizedPeriodicBoundaryConditions[x,y], that parametrize the
twisted periodic boundary conditions across the torus. For the Pfaffian state,
the three groundstates have a combined Chern number of 3.}

    {% Results shown in bandgap figure 
\BothPfaffianBandgapFigure{}
We compute, by exact diagonalization, the low energy band structure of the 3
level JCH system for 4 excitations over the twisted boundary condition manifold
[Fig.~\ref{BothPfaffianBandgapFigure} a)]. For 4 particles there is a large
modulation of the energy as a function of the twist angles (which one expects
to dissapear in the many-particle limit\cite{Varnhagen1995}). We find
that, a quasi-gap can exist in both models and computation of the Chern number
($\ChernNumber = 3$), coupled with the 3-fold degeneracy provides strong
evidence for Pfaffian physics.}
\PfaffianOverlapFigure{}    
{%Interpretation of gap

In the case of Laughlin states on a lattice, the gap has been
shown\cite{Hafezi2007} to scale proportianlly to the flux density per lattice
plaquette. This does not seem to apply in the case of the Pfaffian state. We
hypothesise that the first excited state of this system is not an excitation
lying in a higher Landau level.  Rather, it is some combination of
quasi-particles and quasi-holes. These excitations lie in the LLL, with no
energy gap in the continuum limit, which would explain the rapid decrees in the
gap as the flux density per lattice plaquette decreases.}

{% Results from overlap and gap computations

    {% plots over different lattice sizes and parameters

With strong evidence for the Pfaffian state, we proceed to investigate the
3-level JCH model in more detail, by exploring the properties of the
groundstate over a range of lattice sizes and system parameters.  These
investigations are presented in Fig.~\ref{PfaffianOverlapFigure}. 

In Figs.~\ref{PfaffianOverlapFigure}a) and c) we show how the groundstate
changes as a function of \AtomCavityCoupling[2]. The groundstate experience a
transition from the Pfaffian state away from
$\AtomCavityCoupling[2]/\AtomCavityCoupling[1] = \sqrt{2}$. For
$\AtomCavityCoupling[2]/\AtomCavityCoupling[1] > \sqrt{2}$, where the effective two body interaction becomes
attractive, there is a very sharp transition to a collapsed state. On the other
side, $\AtomCavityCoupling[2]/\AtomCavityCoupling[1] < \sqrt{2}$, the gap and Pfaffian overlap remain fairly
stable, although we observe a transition in the $4\times 5$ lattice
configuration.

In Figs.~\ref{PfaffianOverlapFigure}b) and d) we show how the groundstate changes as
a function of the detuning, $\Delta$. We find that, as in the case of the
Laughlin case\cite{Hayward2012}, increasing the atomic detuning, which alters
the effective interaction strength (Fig. \ref{threeLevelAtomFigure}b), can
induce a transition from a Pfaffian state to an uncorrelated one. This
transition is accompanied by a closing of the bandgap, and, for most cases, a
dropoff of the overlap with the trial wavefunction.  }

The Pfaffian states at $\fillingFactor = 1$ have a straightforward
interpretation as the symmetrised product of two Laughlin states at
$\fillingFactor = 1/2$\cite{Read1996}.  Assigning each particle to one of two Laughlin
states. The wavefunction will vanish as two particles in the same Laughlin state
approach each other, but not if those two particles are in different states.
However, if any three particles coincide, then by construction the wavefunction
will be zero at this point. The three degenerate states in the torus setting
correspond to the singlet and two doublet states one can construct from the
doubly degenerate Laughlin states.

This re-expression of the Pfaffian wavefunction also allows one to translate
findings from investigations into the Laughlin state in the JCH into the
current work. For example, we find that the detuning for which the Pfaffian
state undergoes a transition (Fig.~\ref{PfaffianOverlapFigure}b,d) is the same
for the equivalent single Laughlin state\cite{Hayward2012}. Furthermore, the
overlap with the trial Pfaffian wavefunction is very well approximated by the
overlap with a single Laughlin function, to the power of two. }

{%discussion, and future work

    {%Discussion of Hafiz three body paper
%In \cite{Hafezi2014} Hafezi \etal have shown how the Pfaffian state can be
%realized in a circuit QED setting via an effective 3-body interaction. They
%show the Pfaffian in the hardcore boson limit, where as we study the effect of
%having a multi level atom in the mix. 
%\pdfcomment{Obviously this needs be rewritten in a bit more of a serious way.}
    }

    {%Conclusions

In conclusion, we have described a method by which three-body interactions can
be induced in Jaynes-Cummings-Hubbard systems. In the presence of synthetic
magnetic fields, such interactions, strongly correlated states of light, with
Pfaffian-like topological properties, will exist. This opens up exciting
possibilities for the exploration of exotic quantum states within the cavity
QED framework, including states with non-abelian quasi-particles  pertaining to
topological quantum computing.  } }

{%Acknowledgments 
The authors would like to acknowledge A.D. Greentree for helpful disucssions.
}

\bibliographystyle{apsrev}
\bibliography{\LibraryFile}

\end{document}